\documentclass{moriond}

\def\be{\begin{equation}}
\def\ee{\end{equation}}
\def\bea{\begin{eqnarray}}
\def\eea{\end{eqnarray}}

\newcommand{\Photo}{\includegraphics[height=35mm]{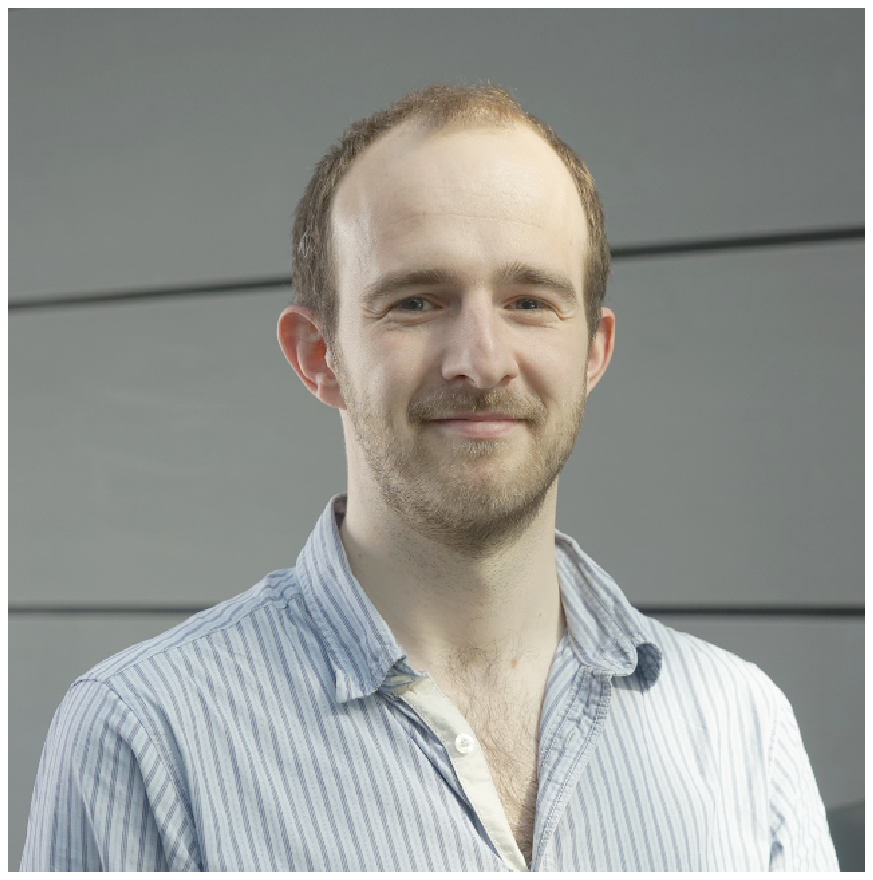}}

\begin{document}
\vspace*{4cm}
\title{One pulsar, two white dwarfs, and a planet confirming the strong equivalence principle}

\author{ G. Voisin }
\address{LUTH, Observatoire de Paris, PSL Research University, CNRS,  5 place Jules Janssen 92195 Meudon, France}
\author{On behalf of: I. Cognard, P.C.C. Freire, N. Wex, L. Guillemot, G. Desvignes, M. Kramer, G. Theureau, M. Saillenfest}

\maketitle\abstracts{
The strong equivalence principle is a cornerstone of general relativity, tested with exquisite accuracy in the Solar system. However, tests in the strong-field regime require a compact object. 
Currently, PSR J0337+1715 is the unique millisecond pulsar found in a triple stellar system, orbiting two white dwarfs within an area comparable to the orbit of the Earth. This configuration offers the opportunity for a dramatic improvement over previous tests, provided that accurate and regular timing of the pulsar can be achieved. This also requires the development of a new timing model solving numerically the relativistic three-body problem with great accuracy. 
We report on the analysis of the high-quality dataset gathered on PSR J0337+1715 by the Nançay radiotelescope over the past 8 years. In particular, I will show how we could obtain the most stringent limit to-date on a potential violation of the strong equivalent principle in the strong field regime. I will also introduce preliminary resuts showing that the presence of a small planet in the system may explain a tiny residual signal so far unaccounted for, which if confirmed would make this system exceptionally rich. }

\section{Introduction}
Tests of the universality of free fall with strongly self-gravitating bodies have strongly benefited from the discovery of the millisecond pulsar PSR J0337+1715 \cite{ransom_millisecond_2014}. Up until that discovery, strong-equivalence-principle (SEP) tests consisted mainly in the Damour-Shafer test\cite{damour_new_1991} whereby one looks for a systematic orbital polarisation of pulsar-white dwarf binaries by the local gravitational potential of the galaxy. This is a kind of three-body test: two test masses, a neutron star and a white dwarf, fall in the gravitational potential of a third one, the Galaxy. If such test suffers primarily from the weakness of the galactic gravitational attraction, it can be partly compensated by the possibility of using a sample of binaries instead of a single system\cite{gonzalez_high-precision_2011}.

PSR J0337+1715 offers a unique opportunity: the system is a hierarchical triple stellar system with two white dwarfs. The innermost white dwarf orbits the pulsar in 1.6 days, while the outermost white dwarf orbits the inner binary in approximately 327 days. Schematically, this configuration allows for a test of the universality of free fall between the two components of the inner binary in the gravitational field of the outer star which is much stronger than the galactic field. In addition, mutual interactions allow for a complete characterisation of the system parameters. For all these reasons, PSR J0337+1715 has permitted a leap forward in the accuracy of SEP tests involving a neutron star \cite{archibald_universality_2018,voisin_improved_2020}.

The principle of the experiment is thus not very different from, for example, Lunar Laser ranging in the Solar system, where Earth and Moon fall in the potential of the Sun. What is massively different is the amount of gravitational energy carried by the neutron star which represents about $\sim-0.15$ of its total mass-energy, against $\sim -10^{-10}$ for the Earth. This allows the test to be sensitive to effects which may occur only in this regime, such as the effect of spontaneous scalarization in the Damour-Esposito Farese scalar-tensor theory\cite{damour_nonperturbative_1993,damour_tensor-scalar_1996}. 

In this contribution, we will present the results obtained from the timing of PSR J0337+1715 at the radiotelescope of Nançay\cite{voisin_improved_2020}(V20 in the following), as well as preliminary results showing hints of the presence of a small planet in the system. 

\section{Testing SEP with PSR J0337+1715}

\subsection{Timing model}
Pulsar timing consists in accurately measuring the times of arrival of pulses from the neutron star. For an isolated star, these are only affected by the spin frequency decay due to the conversion of spin energy into electromagnetic energy and particles, as well as by occasional glitches. When in a binary, the motion of the pulsar on its orbit changes the geometrical path to the observer in a periodic manner. This allows one to extract orbital parameters from the timing data. But since times of arrival are mostly sensitive to the projection of the orbital motion along the line of sight, there remains degeneracies. The first one is between the semi-major axis and the inclination of the orbital plane. The reason is that a Newtonian binary orbit is an ellipse, and the projection of an ellipse remains an ellipse. As a result, one cannot disentangle between all the orbits which share the same projection. The second one results from the fact that a Newtonian binary can be cast into an equivalent single body attracted by a reduced mass which is a function of the two actual masses. In pulsar timing, this is usually shown in terms of the so-called mass function (which is related to, albeit not equal to the reduced mass) which is the observable. Without additional knowledge, masses thus remain degenerate. 

These two symmetries are broken when additional effects are taken into account, in particular the so-called Shapiro and Einstein delays which result from companion-induced time dilation along the path of light and at the source, respectively. Orbital motion can also be affected by relativistic effects such as periastron precession or gravitational-wave emission. With two of these effects, degeneracies can be lifted, and with additional effects general relativity can be put to a consistency test (see P.C.C Freire's contribution on the double pulsar\cite{kramer_strong-field_2021}).

In the case of PSR J0337+1715, these effects are not very strong but detectable except for gravitational wave emissions. Nonetheless, mutual interactions between the three bodies at Newtonian level is in itself sufficient to break the above two symmetries. This comes at the cost of a computationally much more expensive evaluation of the orbital motion. Indeed, contrary to binary pulsars a numerical integration has to be performed with an accuracy of a few meters on the position of the pulsar. This is why we specifically developed a Numerical Timing Model (Nutimo, V20), which solves the equations of motion at first post-Newtonian order and compute the associated delays. A Markov Chain Monte Carlo (MCMC) estimation of the posterior probability density of each parameter of the problem is then performed.

\subsection{Limits on SEP violations}
At Newtonian level, a violation of SEP can be formalised by introducing different gravitational constants for different pairs of interacting bodies. In the case of the triple system, one can show that any measurable violation in the interaction between the two white dwarfs would be larger than limits from Solar system experiments. Thus, we consider that the pulsar interacts with each white dwarf through a constant $G(1+\Delta)$ while they interact with each other through the gravitational constant $G$. 

Thus, $\Delta$ is the SEP violation parameter which is fitted for along with the other parameters of the system. A posterior distribution function is obtained which indicates that $|\Delta| \leq 2\times10^{-6}$ at 95\% confidence level (V20).

To go further, this results needs to be interpreted in the frame of a specific alternative theory of gravity. The parametrised post-Newtonian framework cannot be used because it is not generally valid in the strong field regime, and a Nordvedt parameter cannot be rigorously defined for the same reason\cite{freire_tests_2012}. As an example, we may however cast our result into the frame of the Damour-Esposito Farèse scalar-tensor theory of gravity. This theory depends on two parameters $\alpha_0, \beta_0$ and tends to general relativity when they tend to zero. PSR J0337+1715 is the most constraining observation for a large part of the parameter space (see figure 12 in V20 or 14 in Kramer et al.\cite{kramer_strong-field_2021}), only surpassed by the double pulsar\cite{kramer_strong-field_2021} (see P.C.C. Freire's contribution) and pulsar-white dwarf binaries in the low beta region of the plot. 

\section{A small planet perturbing the system ?}

\subsection{Hint: a low frequency signal in the fit residuals}
The fit residuals of V20 exhibited low-frequency noise which we initially associated with red noise, a phenomenon quite common in pulsar timing (see e.g. A. Chalumeau's contribution\cite{chalumeau_noise_2022}). Red noise can have either intrinsic causes, presumably due to variations in the emission region of the pulsar magnetosphere, or be caused by propagation effects. The latter results from variations of the column density of free electrons along the line of sight, the so-called dispersion measure, and is a chromatic effect, contrary to the former. In the present case, no sign of chromaticity is present, and it can be ruled out. 

On the other hand, the small amplitude of noise, about $1\rm\mu s$, and its timescale of a few thousand days are compatible with a small planet orbiting the stellar triple system at a sufficiently large distance such that its orbit may remain stable. Although rare, planets have been found around pulsars (see e.g. Ni\c tu et al.\cite{nitu_search_2022} and references therein). We therefore implemented a fourth body in Nutimo in order to test that hypothesis.

\subsection{Preliminary results}
\begin{figure}
	\caption{Lomb-Scargle periodogram of fit residuals of the planet model. The three grey horizontal lines represent the mean and the upper and lower limit of the 99\% region obtained by bootstrapping white noise. These lines have been smoothed by a sliding mean over 40 frequency bins to improve clarity. $f_I, f_O, f_E$ are the inner binary, outer white dwarf and Earth orbital frequencies respectively. $2f_I-f_O$ is the frequency of the signal produced by SEP violation. \label{fig:LS}  }
	\begin{center}
		\includegraphics[width=\textwidth]{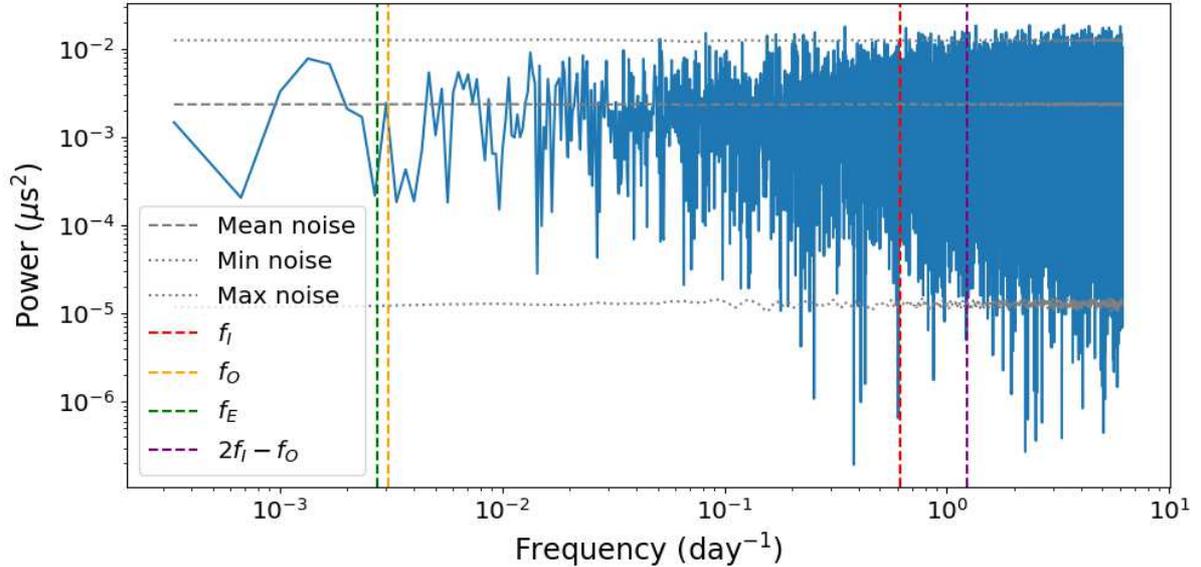}
	\end{center}
\end{figure}
The planet model is able to absorb all the low-frequency components down to the point where residuals show a flat periodogram compatible with white noise, Fig. \ref{fig:LS}. The planet has a mass of a few $10^{-8} M_\odot$, comparable to the mass of the moon and an orbital period of $\sim 3000$ days. When compared to V20, the best fit has a significantly better $\chi^2$ even when a penalty against additional parameters is applied using the Akaike Information Criterion (AIC) or the Bayesian information criterion (BIC), as shown in table \ref{tab:comp}.

\begin{table}[t]
	\caption[]{Comparison between the planet model and the results of V20. $N_{\rm par}$ is the number of parameters of the model and $N_{\rm dof} = 9303 - N_{\rm par}$ is the number of degrees of freedom. }
	\label{tab:comp}
	\vspace{0.4cm}
	\begin{center}
		\begin{tabular}{|cccccc|}
			\hline
			\\
			Model & $N_{\rm par}$ & $\chi^2$ & $\chi^2/N_{\rm dof}$ & $\Delta \rm AIC$ & $\Delta \rm BIC$ \\
			\hline
			V20 & 25 & 16066 & 1.7316 & 0 & 0 \\
			Planet & 32 & 13185 & 1.4220 & -2869 & -2827 \\
			\hline
		\end{tabular}
	\end{center}
\end{table}

In V20, the low-frequency signal was accounted for by rescaling uncertainties on times of arrival such that the reduced $\chi^2$ would be unity. Although the frequency of a SEP violation would be much higher \cite{archibald_universality_2018}, and therefore is not expected to be biased by such low-frequency components, this procedure allowed us to produce a conservative estimate of the uncertainty on the SEP violation parameter $\Delta$ (as well as on every other parameter). Conversely, the presence of the planet does not affect directly the SEP test, but the rescaling of error bars is more limited, which translates into a narrower uncertainty on every parameter, including $\Delta$.  

\section{Conclusion}
 We have shown that the timing of PSR J0337+1715, although complex to model and computationally challenging, provides one of the most constraining tests of the strong equivalence principle to date. Preliminary results show that the addition of a small planet to the model allows for a significantly better fit to the data which leads to an improved limit on a putative violation of the equivalence principle. If confirmed, this would also make the formation channel of this system particularly intriguing. 

%
%
%

\section*{References}

\bibliography{voisin}
%
%
%
%

\end{document}